\documentclass[aps,pre,twocolumn,nofootinbib]{revtex4}

\usepackage{graphicx,amssymb,amsfonts,amsmath,chemarr,color,tcolorbox,float,ragged2e}

\newcommand{\beq}{\begin{equation}}
\newcommand{\eeq}{\end{equation}}
\newcommand{\beqn}{\begin{eqnarray}}
\newcommand{\eeqn}{\end{eqnarray}}
\newcommand{\elabel}[1]{\label{eq:#1}}
\newcommand{\eref}[1]{Eqn.\ \ref{eq:#1}}
\newcommand{\erefs}[2]{Eqns.\ \ref{eq:#1} and \ref{eq:#2}}

\newcommand{\erefn}[2]{Eqns.\ \ref{eq:#1}-\ref{eq:#2}}
\newcommand{\flabel}[1]{\label{fig:#1}}
\newcommand{\fref}[1]{Fig.\ \ref{fig:#1}}

\newcommand{\slabel}[1]{\label{sec:#1}}
\newcommand{\sref}[1]{Sec.\ \ref{sec:#1}}
\newcommand{\alabel}[1]{\label{app:#1}}
\newcommand{\aref}[1]{Appendix \ref{app:#1}}
\newcommand{\ket}[1]{|#1\rangle}

\newcommand{\avg}[1]{\langle#1\rangle}

\newcommand{\Lh}{\hat{\cal L}}
\renewcommand{\a}{\hat{a}}
\newcommand{\ad}{\hat{a}^\dagger}
\renewcommand{\b}{\hat{b}}
\newcommand{\bd}{\hat{b}^\dagger}

\newfloat{mybox}{tb}

\begin{document}

\title{Stochastic modeling of gene expression, protein modification, and polymerization}

\author{Andrew Mugler}
\email{amugler@purdue.edu}
\affiliation{Department of Physics and Astronomy, Purdue University, West Lafayette, IN 47907, USA}

\author{Sean Fancher}
\affiliation{Department of Physics and Astronomy, Purdue University, West Lafayette, IN 47907, USA}

\begin{abstract}
Many fundamental cellular processes involve small numbers of molecules. When numbers are small, fluctuations dominate, and stochastic models, which account for these fluctuations, are required. In this chapter, we describe minimal stochastic models of three fundamental cellular processes: gene expression, protein modification, and polymerization. We introduce key analytic tools for solving each model, including the generating function, eigenfunction expansion, and operator methods, and we discuss how these tools are extended to more complicated models. These analytic tools provide an elegant, efficient, and often insightful alternative to stochastic simulation.
\end{abstract}

\maketitle

Cells perform complex functions using networks of interacting molecules, including DNA, mRNA, and proteins. Many of these molecules are present in very low numbers per cell. For example, over 80\% of the genes in the {\it E.\ coli} bacterium express fewer than a hundred copies of each of their proteins per cell \cite{guptasarma1995}. When the numbers are this small, fluctuations in these numbers are large. Indeed, we will see in this chapter that the simplest model of gene expression predicts Poisson statistics, meaning that the standard deviation equals the square root of the mean. For means of $100$, $10$, and $1$ proteins, fluctuations are $10\%$, $32\%$, and $100\%$ of the mean, respectively. Most manmade devices would not function properly with fluctuations this large. But for a cell these fluctuations are unavoidable: they are not due to external factors, but rather they arise intrinsically due the small numbers. Experiments in recent years have vividly demonstrated that number fluctuations are ubiquitous in microbial and mammalian cells alike \cite{elowitz2002, raj2006} and occur even when external factors are held constant \cite{elowitz2002}.

From a mathematical modeling perspective, accounting for large fluctuations requires models that describe not just the mean molecule numbers, but rather the full distributions of molecule numbers. These are stochastic models. By far the most common way to solve stochastic models has been by computer simulation \cite{gillespie1977, erban2007}. Typically one simulates many fluctuating trajectories of molecule numbers over time, and then builds from these trajectories the molecule number distribution. This technique can be applied to arbitrarily complex reaction networks and provides exact results in the limit of infinite simulation data. However, simulations can be inefficient (although faster approximate schemes have been developed in recent years \cite{gillespie2001, rathinam2003}), and, perhaps more importantly, simulations do not readily provide the physical intuition that analytic solutions provide. Therefore, many researchers have devoted attention to developing methods for obtaining exact or approximate analytic solutions to stochastic models \cite{hornos2005, shahrezaei2008, iyerbiswas2009, walczak2009, mugler2009, kumar2014, hespanha2005, munsky2006}.

In this chapter, we describe minimal stochastic models of three fundamental cellular processes: gene expression, protein modification, and polymerization. All are exactly solvable, and our focus here is on introducing the key analytic tools that can be used to solve them and gain physical insight about their behavior. These tools include the use of a generating function, the expansion of distributions in their natural eigenfunctions, and the use of operator methods originally derived from quantum mechanics. The goal is to provide readers with these tools so that they may see how to apply them to new stochastic problems. To that end, we conclude the chapter with a discussion of how these tools have seen recent application to models of more complex phenomena, including gene regulation, cell signaling networks, and more detailed mechanisms of polymer growth.

With the exception of new results for the polymerization model (\sref{poly}), this chapter is a review. The generating function is a canonical tool that is discussed in several classic textbooks on stochastic processes \cite{vankampen1992, gardiner1985}. The use of quantum operator methods in a biochemical context dates back to the 1970s \cite{doi1976, zeldovich1978, peliti1986} and has been nicely reviewed \cite{mattis1998}. The use of eigenfunctions to solve stochastic equations has been recently developed in the contexts of gene regulation \cite{walczak2009, mugler2009, walczak2012} and spatially distributed cell signaling \cite{mugler2013}. Thus, the aim of this chapter is to provide a unified and accessible introduction to all of these tools, using three fundamental processes from cell biology.

\begin{figure*}
\centering
\includegraphics[width = \textwidth]{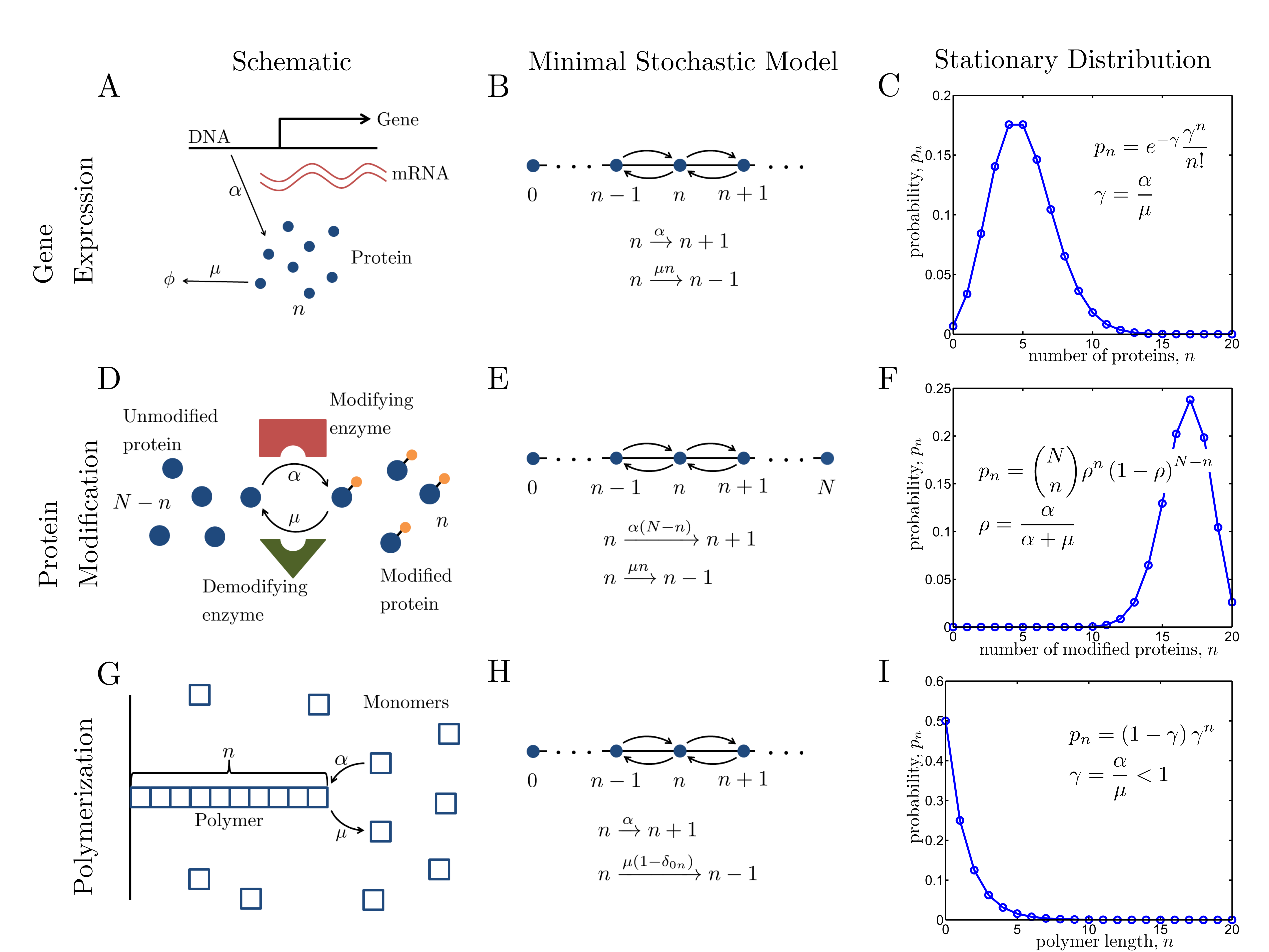}
\caption{Stochastic modeling of gene expression, protein modification, and polymerization. (A-C) In a minimal model of gene expression, proteins are produced at rate $\alpha$ and removed at rate $\mu$; there are $n$ proteins at any given time. The probability $p_n$ is a Poisson distribution in steady state. Different minimal models describe (D-F) protein modification by enzymes, where the steady state is a binomial distribution, and (G-I) polymerization, where the steady state is a geometric distribution. The parameters are $\gamma = 5$ (C), $\rho = 5/6$ and $N = 20$ (F), and $\gamma = 1/2$ (I).}
\flabel{processes}
\end{figure*}

\section{Gene expression}

We begin with a discussion of gene expression, which is the process of producing proteins from DNA. As depicted in \fref{processes}A, a particular segment of the DNA (the gene) is transcribed into mRNA molecules, which are then translated into proteins. The processes of transcription and translation can be highly complex, especially in higher organisms, but for the purposes of minimal modeling we omit these details and refer the reader to several excellent sources for more information \cite{alberts2007, phillips2012}. Typically mRNAs are degraded with a timescale on the order of minutes, whereas proteins are removed from the cell (either via degradation or dilution from cell division) with a timescale on the  order of tens of minutes to hours \cite{alon2006}. This timescale separation allows us, in a minimal model, to approximate the mRNA number as roughly constant in time and focus on the protein number $n$ as our only degree of freedom. This model of gene expression neglects such common features such as regulated protein production and the production of proteins in bursts, both of which are further discussed in \sref{extensions}.

The stochastic model of gene expression is given by the master equation, which specifies the dynamics of the probability $p_n$ of having $n\in\{0,1,2,\dots,\infty\}$ proteins per cell. Introducing $\alpha$ as the rate of protein production and $\mu$ as the rate of protein removal, the master equation reads
\beq
\elabel{me_gene}
\frac{dp_n}{dt} = \alpha p_{n-1} + \mu (n+1) p_{n+1} - \alpha p_n - \mu n p_n.
\eeq
The four terms on the right-hand side reflect the four ways of either entering or leaving the state with $n$ proteins (\fref{processes}B). Production occurs with a constant propensity $\alpha$, whereas removal occurs with propensity $\mu n$, since any of the $n$ proteins has a chance of being removed. \eref{me_gene} is modified at the $n=0$ boundary: the first and fourth terms are absent since transitions from and to the $n=-1$ state are prohibited, respectively. \eref{me_gene} is often called the birth-death process.

The birth-death process admits a steady-state (or stationary) solution, where $dp_n/dt = 0$. By considering the equations for $n=0, 1, 2, \dots$ in succession, one readily notices a pattern (see \aref{exercises}). The result is
\beq
\elabel{poisson}
p_n = e^{-\gamma}\frac{\gamma^n}{n!},
\eeq
where $\gamma \equiv \alpha/\mu$. \eref{poisson} is the Poisson distribution (\fref{processes}C). As we will show below, it has the property that its mean $\avg{n} = \sum_{n=0}^\infty p_n n$ equals its variance $\sigma^2 = \avg{n^2} - \avg{n}^2$. Therefore, relative fluctuations go down with the mean, $\sigma/\avg{n} = \avg{n}^{-1/2}$, which explains why small mean protein numbers correspond to large relative fluctuations.

\subsection{The generating function}

We now introduce the generating function, which is a highly useful tool for solving stochastic equations. The generating function is defined by
\beq
\elabel{Gdef}
G(z) = \sum_n p_n z^n
\eeq
for some continuous variable $z$. Its name comes from the fact that moments of $p_n$ are {\it generated} by derivatives of $G(z)$ evaluated at $z=1$,
\beqn
\elabel{Gnorm}
G(1) &=& \sum_n p_n = 1, \\
\elabel{Gmean}
G'(1) &=& \sum_n p_n n = \avg{n}, \\
\elabel{Gsecond}
G''(1) &=& \sum_n p_n n(n-1) = \avg{n^2} - \avg{n},
\eeqn
and so on. We invert the relationship in \eref{Gdef}, also by taking derivatives, but evaluating at $z=0$,
\beq
\elabel{pdef}
p_n = \frac{1}{n!}\partial_z^n\left[G(z)\right]_{z=0}.
\eeq
\eref{pdef} is verified by inserting \eref{Gdef} and recognizing that $\lim_{z\to0} z^m = \delta_{m0}$.

The generating function greatly simplifies the master equation by turning a set of coupled ordinary differential equations (one for each value of $n$ in \eref{me_gene}) into a single partial differential equation. For the birth death-process, we derive this partial differential equation by multiplying \eref{me_gene} by $z^n$ and summing both sides over $n$ from $0$ to $\infty$ (see \aref{exercises}). The result is
\beq
\elabel{pde_gene}
\partial_t G = -(z-1)(\mu\partial_z - \alpha)G,
\eeq
where the appearances of $z$ and $\partial_z$ are due to the shifts $n-1$ and $n+1$, respectively. \eref{pde_gene} is readily solved in steady state, where $\partial_t G = 0$. There we must have $\mu \partial_zG = \alpha G$, and thus
\beq
\elabel{Ggene}
G(z) = e^{-\gamma}e^{\gamma z},
\eeq
where once again $\gamma = \alpha/\mu$, and here the factor of $e^{-\gamma}$ follows from the normalization condition in \eref{Gnorm}. Repeatedly differentiating \eref{Ggene} according to \eref{pdef} immediately gives the Poisson distribution, \eref{poisson}. Furthermore, differentiating \eref{Ggene} according to \erefs{Gmean}{Gsecond} gives $\sigma^2 = \avg{n} = \gamma$, confirming the relationship between the variance and the mean.

The full time-dependent solution of \eref{pde_gene} is obtained either by applying the method of characteristics \cite{walczak2012} or by a transformation of variables \cite{gardiner1985}. We present the latter method here. Writing $G(z,t) = H(z,t)e^{\gamma(z-1)}$ transforms \eref{pde_gene} into
\beq
\elabel{pde_H}
\partial_t H = -\mu(z-1)\partial_z H,
\eeq
and writing $z - 1 = e^y$ transforms \eref{pde_H} into
\beq
\elabel{pde_y}
\partial_t H = -\mu\partial_y H.
\eeq
\eref{pde_y} is a first-order wave equation, whose solution is any function of $y-\mu t$. For convenience we write this function as $H(z,t) = F(e^{y-\mu t}) = F[(z-1)e^{-\mu t}]$, such that
\beq
\elabel{Ggene_t}
G(z,t) = F[(z-1)e^{-\mu t}]e^{\gamma(z-1)}.
\eeq
The unknown function $F$ is determined by the initial condition \cite{gardiner1985}. Note that normalization (\eref{Gnorm}) requires $G(1,t) = F(0) = 1$, which confirms that $G(z,t\to\infty) = e^{\gamma(z-1)}$ in steady state, as in \eref{Ggene}.

\begin{figure*}
\centering
\includegraphics[width = \textwidth]{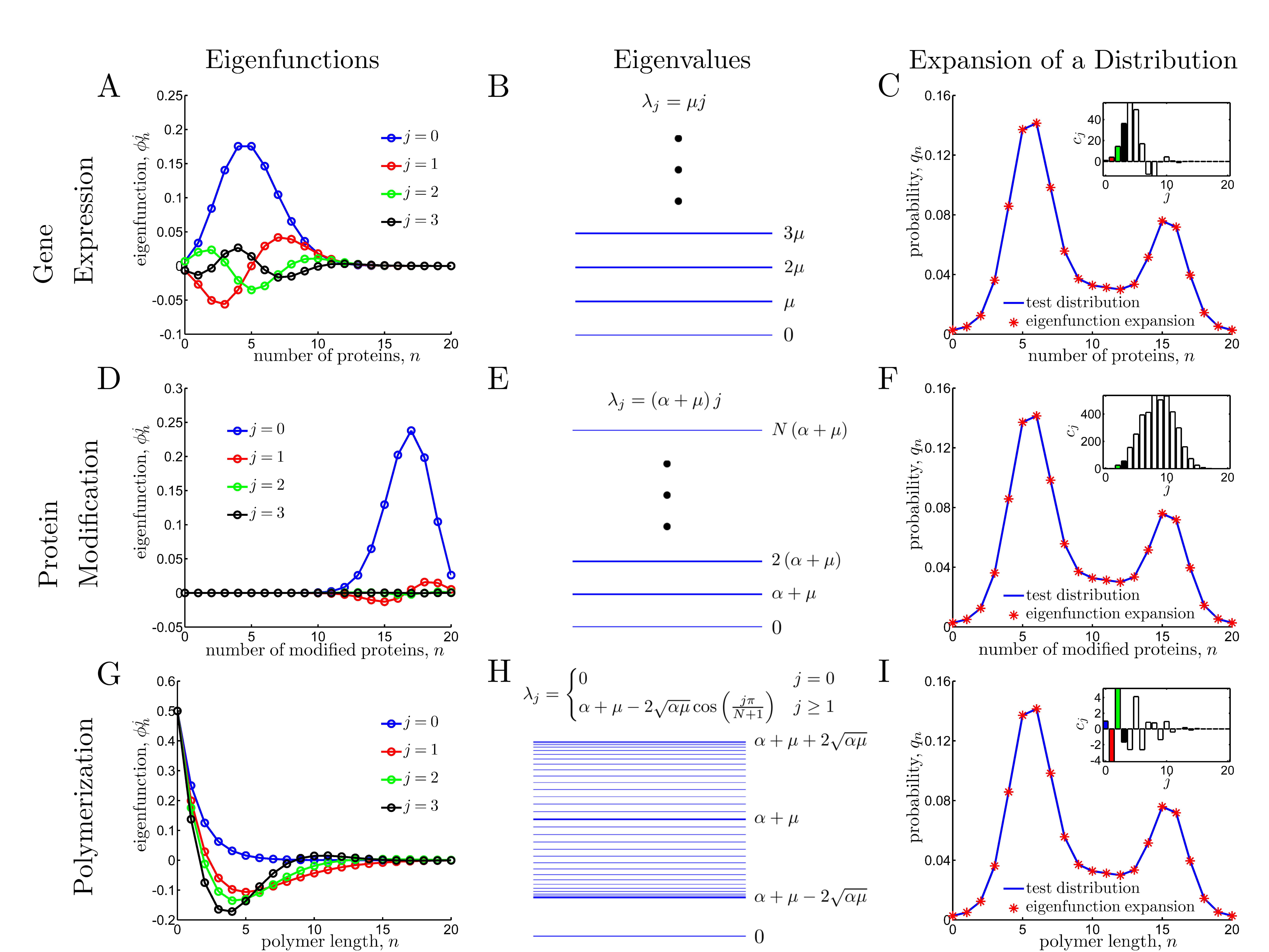}
\caption{Eigenfunctions and eigenvalues of the gene expression, protein modification, and polymerization models. During the time evolution of each stochastic process, the eigenfunctions (A, D, G) relax according to rates given by the eigenvalues (B, E, H). The ``zero mode'' $\phi_n^0$ is therefore the steady state distribution. In all three cases, the eigenfunctions also form a basis in which any distribution $q_n$ can be expanded (C, F, I), which facilitates analytic solutions; the insets give the expansion coefficients $c_j$. The parameters are $\gamma = 5$ (A), $\rho = 5/6$ and $N=20$ (D), and $\gamma = 1/2$ and $N=20$ (G). In C, F, and I, the parameters $\gamma$, $\rho$, and $\gamma$, respectively, act as ``gauge freedoms'', since they affect $c_j$ but not the reconstruction of $q_n$; they are set to $\gamma = 5$ (C), $\rho = 3/7$ (F), and $\gamma = 1/2$ (I).}
\flabel{eigen}
\end{figure*}

\subsection{Eigenvalues and eigenfunctions}

The master equation is a linear equation. That is, \eref{me_gene} is linear in $p$, and \eref{pde_gene} is linear in $G$. This means that the master equation is conveniently solved by expanding in the eigenfunctions of its linear operator. We will see that exploiting the eigenfunctions not only provides an alternative to the solution techniques presented thus far, but that the eigenfunctions are useful in their own right. They provide insights on the dynamics, they form a complete basis in which any probability distribution can be expanded, and they facilitate extension to more complex models of gene expression and regulation.

The linear operator for the birth-death process is evident from \eref{pde_gene}. Writing \eref{pde_gene} as $\partial_t G = -\Lh G$, we see that $\Lh = (z-1)(\mu\partial_z-\alpha)$. The eigenfunctions of $\Lh$ then satisfy
\beq
\elabel{eigen}
\Lh\phi_j(z) = \lambda_j \phi_j(z)
\eeq
for eigenvalues $\lambda_j$. Inserting the form for $\Lh$, we see that \eref{eigen} is a first-order ordinary differential equation for $\phi_j(z)$ that can be solved by separating variables and integrating. The result is
\beq
\elabel{ef_gene}
\phi_j(z) = (z-1)^{\lambda_j/\mu}e^{\gamma(z-1)}
\eeq
up to a constant prefactor. We set the prefactor to one by recognizing that for $\lambda_j=0$, \eref{eigen} is equivalent to the master equation in steady state, and therefore \eref{ef_gene} should recover the steady-state solution (\eref{Ggene}) when $\lambda_j = 0$. As in \eref{pdef}, the eigenfunctions are converted into $n$ space via
\beq
\elabel{phidef}
\phi_n^j = \frac{1}{n!}\partial_z^n[\phi_j(z)]_{z=0}.
\eeq
Example eigenfunctions are shown in \fref{eigen}A. Note that for $\lambda_j=0$, the eigenfunction is the Poissonian steady-state distribution, \eref{poisson}.

We now demonstrate the solution of \eref{pde_gene} by eigenfunction expansion. We expand
\beq
\elabel{Gexp}
G(z,t) = \sum_j C_j(t) \phi_j(z)
\eeq
and insert it into \eref{pde_gene} to obtain
\beq
\sum_j \left(\partial_t C_j\right) \phi_j = -\Lh \sum_j C_j \phi_j = \sum_j \left(-\lambda_j C_j\right) \phi_j,
\eeq
where the second step follows from the eigenvalue relation, \eref{eigen}.
Equating the terms in parentheses for each $j$, we obtain an ordinary differential equation that is solved by
$C_j(t) = c_je^{-\lambda_j t}$ for initial conditions $c_j$. Inserting this form and \eref{ef_gene} into \eref{Gexp}, we find
\beq
\elabel{Geigen}
G(z,t) = e^{\gamma(z-1)}\sum_j c_j e^{-\lambda_j t} (z-1)^{\lambda_j/\mu}.
\eeq
This expression can be directly compared with our previous solution, \eref{Ggene_t}, by Taylor expanding $F$ as $F(x) = \sum_{j=0}^\infty x^j\partial_x^j [F(x)]_{x=0}/j!$. Then \eref{Ggene_t} becomes
\beqn
G(z,t) &=& \sum_{j=0}^\infty \left[ (z-1)e^{-\mu t} \right]^j \frac{\partial_x^j[F(x)]_{x=0}}{j!} e^{\gamma (z-1)} \nonumber \\
\elabel{Geigen2}
&=& e^{\gamma(z-1)}\sum_{j=0}^\infty \frac{\partial_x^j[F(x)]_{x=0}}{j!} e^{-j\mu t} (z-1)^j. \qquad
\eeqn
Comparing \eref{Geigen2} with \eref{Geigen} term by term, we conclude that
\beq
\elabel{ev_gene}
\lambda_j = \mu j,
\eeq
for $j \in \{0,1,2,\dots,\infty\}$. \eref{ev_gene} gives the eigenvalues of the birth-death process, as depicted in \fref{eigen}B. We also conclude that $F$ and $c$, which are both determined by the initial condition, must be related by $c_j = \partial_x^j [F(x)]_{x=0}/j!$.

\eref{Geigen} shows that the eigenvalues dictate the dynamics of the time-dependent solution. That is, the solution is built from a linear combination of the eigenfunctions, each of which decays exponentially with time, and the eigenvalues set the rates of decay. Larger eigenvalues correspond to faster rates of decay, and in the end there is only one eigenfunction left: the ``zero mode'', with eigenvalue $\lambda_0 = 0$. Hence, the zero mode is the steady state.

The linear operator $\Lh$ is not Hermitian. This means that the eigenfunctions $\phi_n^j$ are not orthogonal to one another. Instead, a different set of conjugate eigenfunctions $\psi_n^j$ is required to satisfy the orthonormality relation $\sum_n \phi_n^j \psi_n^{j'} = \delta_{jj'}$. Now that we know the eigenvalues, we obtain the eigenfunctions in $n$ space by inserting \eref{ef_gene} into \eref{phidef} and evaluating the derivatives (see \aref{exercises}). The result is
\beq
\elabel{phi_gene}
\phi_n^j = e^{-\gamma}\frac{\gamma^n}{n!} \sum_{\ell=0}^{\min(n,j)} (-1)^{j-\ell} {n\choose\ell} {j\choose\ell}
	\frac{\ell!}{\gamma^\ell}.
\eeq
Each eigenfunction is the Poisson distribution multiplied by a $j$th-order polynomial in $n$. In fact, each eigenfunction is the negative discrete derivative of the previous one, $\phi_n^{j+1} = -( \phi_n^j - \phi_{n-1}^j )$ \cite{walczak2012}, which is evident from \fref{eigen}A. Given \eref{phi_gene}, the conjugate eigenfunctions $\psi_n^j$ are constructed to obey the orthonormality relation \cite{walczak2012}. They read
\beq
\elabel{psi_gene}
\psi_n^j = \frac{\gamma^j}{j!} \sum_{\ell=0}^{\min(n,j)} (-1)^{j-\ell} {n\choose\ell} {j\choose\ell}
	\frac{\ell!}{\gamma^\ell}.
\eeq
They are $j$th-order polynomials in $n$.

Together, \erefs{phi_gene}{psi_gene} form a complete basis in which any arbitrary probability distribution $q_n$ can be expanded \cite{mugler_thesis}, as demonstrated in \fref{eigen}C. Explicitly, we write
\beq
\elabel{qdef}
q_n = \sum_{j=0}^\infty c_j \phi_n^j,
\eeq
where the coefficients $c_j$ are the projections of $q_n$ against the conjugate eigenfunctions,
\beq
\elabel{cdef}
c_j = \sum_{n=0}^\infty q_n \psi_n^j.
\eeq
Even though an infinite number of eigenfunctions are needed to complete the expansion, in most practical cases the coefficients $c_j$ die off as a function of $j$ (\fref{eigen}C inset), allowing one to truncate the sums in \erefs{qdef}{cdef} according to the desired numerical accuracy. Moreover, in this context $\gamma$ acts as a free parameter for the expansion, much like a gauge freedom in field theory, and can be tuned to minimize numerical error. The completeness property allows one to expand more complex models of gene regulation in these simpler birth-death eigenfunctions, as further discussed in \sref{extensions}.

\subsection{Operator methods}

The master equation can also be recast in a form that uses raising and lowering operators, familiar to physicists from the operator treatment of the quantum harmonic oscillator \cite{townsend2000}. The idea, detailed in the box, is that these operators raise and lower the protein number by one, and analogous operators raise and lower the eigenvalues by one as well. The operators provide an elegant way to perform linear algebraic manipulations, and they facilitate extension to more complex models \cite{walczak2009, mugler2009}. They also allow one to show that many useful properties, including orthonormality and completeness of the eigenfunctions, are inherited from the Hermitian quantum problem \cite{mugler_thesis}.

\begin{mybox}[tb]
\begin{tcolorbox}
\justifying
\begin{center}
{\bf Raising and lowering operators}
\end{center}
In the operator notation, the generating function is introduced as an expansion over a complete set of abstract states, indexed by $n$,
\beq
\ket{G} = \sum_n p_n \ket{n}. \nonumber
\eeq
The dynamics of $\ket{G}$ are obtained by summing the master equation against $\ket{n}$. For the birth-death process (\eref{me_gene}), we find, similar to \eref{pde_gene},
\beq
\partial_t\ket{G} = -(\ad-1)(\mu\a-\alpha)\ket{G}, \nonumber
\eeq
where $\ad$ and $\a$ are raising and lowering operators. Just as in quantum mechanics (but with slightly different prefactors), they obey
\beq
\ad\ket{n} = \ket{n+1}, \qquad \a\ket{n} = n\ket{n-1}. \nonumber
\eeq
They satisfy the familiar commutation relation, and $\ad\a$ acts as the number operator,
\beq
[\a,\ad] = 1, \qquad \ad\a\ket{n} = n\ket{n}. \nonumber
\eeq
The dynamics of $\ket{G}$ can be written $\partial_t\ket{G} = -\mu\bd\b\ket{G}$ if we define
\beq
\bd = \ad -1, \qquad \b = \a-\gamma, \nonumber
\eeq
where $\gamma = \alpha/\mu$. The operators $\bd$ and $\b$ are raising and lowering operators as well, not for the $\ket{n}$ states, but for the eigenstates $\ket{j}$,
\beq
\bd\ket{j} = \ket{j+1}, \qquad \b\ket{j} = j\ket{j-1}. \nonumber
\eeq
Since $\bd\b$ is also a number operator, it is clear that the eigenvalues of $\Lh = \mu \bd\b$ are $\lambda_j = \mu j$, just as in \eref{ev_gene}. For more details on operator methods as applied to stochastic problems, see \cite{mattis1998} for a general review and \cite{walczak2012} for applications to gene expression.
\end{tcolorbox}
\end{mybox}

\section{Protein modification}

Cells respond to signals in their environment on timescales faster than the minutes to hours required for proteins to be produced. They do so by modifying proteins that are already present, for example by adding or removing phosphate groups (see \fref{processes}D). Modification is typically performed by enzymes such as kinases and phosphatases, and can occur on timescales of seconds or fractions of a second \cite{lim2014}. This makes protein modification much faster than protein production.

A minimal stochastic model of protein modification therefore assumes that the total number of proteins $N$ in the cell or cellular compartment is approximately constant on the timescale of modification. The degree of freedom $n$ is then the number of modified proteins, and $N-n$ is the number of unmodified proteins. Calling $\alpha$ and $\mu$ the modification and demodification rates, the master equation reads
\beqn
\frac{dp_n}{dt} &=& \alpha [N-(n-1)]p_{n-1} + \mu (n+1) p_{n+1} \nonumber \\
\elabel{me_mod}
	&&- \alpha (N-n) p_n - \mu n p_n.
\eeqn
This equation is similar to that for gene expression (\eref{me_gene}) but with two important differences: (i) $n$ is bounded from both sides, $n \in \{0, 1, 2, \dots, N\}$, and (ii) the modification propensity $\alpha (N-n)$ is not a constant, but rather it depends on the number $N-n$ of unmodified proteins available for modification (\fref{processes}E). At the $n=0$ boundary the first and fourth terms in \eref{me_mod} are absent, while at the $n=N$ boundary the second and third terms are absent.

The steady state of \eref{me_mod} is readily found by iteration and pattern matching,
\beq
\elabel{binomial}
p_n = {N\choose n} \rho^n (1-\rho)^{N-n},
\eeq
where $\rho \equiv \alpha/(\alpha + \mu)$. \eref{binomial} is the binomial distribution (\fref{processes}F). It emerges as the steady state because it describes the probability of achieving $n$ successes out of a total of $N$ binary trials, where $\rho$ is the success probability. Here, the trial is whether or not a given protein is modified, and $\rho$ is the modification probability. The equation for the generating function can be derived in the same way as above and reads
\beq
\elabel{pde_mod}
\partial_t G = -(z-1)[(\alpha z+\mu)\partial_z - \alpha N]G.
\eeq
The steady state is
\beq
G(z) = \left[ \rho(z-1) + 1 \right]^N,
\eeq
which recovers \eref{binomial} when repeatedly differentiated according to \eref{pdef}.

The eigenvalues and eigenfunctions of the protein modification process are derived in much the same way as for gene expression above \cite{mugler2013}. The eigenvalues are
\beq
\lambda_j = (\alpha+\mu)j,
\eeq
for $j \in \{0,1,2,\dots,N\}$. Unlike for gene expression, the eigenvalues here depend on both $\alpha$ and $\mu$ because both the modification and demodification propensities are linear in $n$. This means that both $\alpha$ and $\mu$ determine the relaxation dynamics, instead of just $\mu$. The eigenfunctions are given in $z$ space by
\beq
\elabel{ef_mod}
\phi_j(z) = \left[(1-\rho)(z-1)\right]^j \left[ \rho(z-1) + 1 \right]^{N-j},
\eeq
and in $n$ space by
\beq
\elabel{phi_mod}
\phi_n^j = \sum_{\ell \in \Omega} (-1)^{j-n+\ell} {N-j\choose \ell} {j\choose n-\ell} \rho^\ell (1-\rho)^{N-\ell},
\eeq
where $\Omega$ is defined by $\max(0,n-j)\le\ell\le\min(n,N-j)$. The eigenfunctions and eigenvalues are shown in \fref{eigen}D and E. Just as with gene expression, any distribution $q_n$ can be expanded in the eigenfunctions, as shown in \fref{eigen}F. Here $\rho$ acts as the free parameter. The expansion follows \erefs{qdef}{cdef}, with the conjugate eigenfunctions given by \cite{mugler2013}
\beq
\psi_n^j = \frac{1}{(1-\rho)^j}\sum_{\ell\in\Omega} {N-j+\ell\choose\ell} {n\choose j-\ell} (-\rho)^\ell,
\eeq
where $\Omega$ is defined by $\max(0,j-n)\le\ell\le j$. In fact, no truncation is necessary in this case since $n$, and thus $j$, is explicitly bounded between $0$ and $N$.

\section{Polymerization}
\slabel{poly}

One of the major functions of proteins is to provide cells with mechanical capabilities. For example, cell rigidity and mobility are provided by networks of polymers, such as microtubules and actin filaments \cite{boal2012}. A polymer is a linear chain of monomer proteins that attach to and detach from the polymer at one or both ends. The attachment and detachment processes make polymers highly dynamic objects that often undergo rapid and appreciable length fluctuations over a cell's lifetime.

Here we consider a minimal stochastic model of a polymer that changes dynamically at one end only (\fref{processes}G). In this case, the degree of freedom $n \in \{0,1,2,\dots,\infty\}$ is the number of monomers in the polymer, and $\alpha$ and $\mu$ define the attachment and detachment rates. The master equation reads
\beq
\elabel{me_poly}
\partial_t p_n = \alpha p_{n-1} + \mu p_{n+1} - \alpha p_n - \mu p_n.
\eeq
This equation differs from the previous two examples (\erefs{me_gene}{me_mod}) in that neither the attachment nor the detachment propensity is linear in $n$ (\fref{processes}H). This is because both attachment and detachment occur only at the polymer tip, and so neither process is influenced by how many monomers are already part of the polymer. The important exception is the case when $n=0$; here we must force the detachment propensity to be zero, since there are no actual monomers to detach. This accounts for the highly nonlinear detachment propensity $\mu(1-\delta_{n0})$ shown in \fref{processes}H, and implies that at the $n=0$ boundary in \eref{me_poly} the first and fourth terms are absent.

The steady state of \eref{me_poly} is once again found by iteration and pattern matching,
\beq
\elabel{geometric}
p_n = (1-\gamma)\gamma^n
\eeq
where $\gamma \equiv \alpha/\mu$. We see that we must have $\alpha < \mu$ for \eref{geometric} to be valid. This is because in the opposite regime $\alpha > \mu$, attachment outpaces detachment, and the polymer length diverges. Therefore we restrict ourselves here to the regime $\alpha < \mu$, where detachment dominates, and the polymer length distribution has a non-divergent steady state. \eref{geometric} is the geometric distribution, which is the discrete analog of the exponential distribution. It is illustrated in \fref{processes}I.

The dynamics of the generating function obey
\beq
\elabel{pde_poly}
\partial_t G = -(z-1)\left(\frac{\mu}{z}-\alpha\right)G + (z-1)\frac{\mu}{z}p_0(t).
\eeq
Note that \eref{pde_poly} is not a partial differential equation as in the previous two cases. Instead, it is a non-homogeneous ordinary differential equation in time, where the forcing term is proportional to the unknown dynamic function $p_0(t)$. In steady state, this function is a constant $p_0$, which is set by the normalization condition $G(1) = 1$, yielding
\beq
\elabel{Gpoly}
G(z) = \frac{1-\gamma}{1-\gamma z}.
\eeq
As expected, \eref{Gpoly} recovers \eref{geometric} when repeatedly differentiated according to \eref{pdef}.

The presence of the $p_0(t)$ term in \eref{pde_poly} makes it more difficult than in the previous two cases to find the eigenvalues and eigenfunctions using the generating function. Nonetheless, since the problem is still perfectly linear in $p$, we make progress directly in $n$ space. To do so, we write \eref{me_poly} as a matrix equation, $\partial_t \vec{p} = -{\cal L} \vec{p}$, where
\beq
\elabel{tridiag}
{\cal L} =
\begin{pmatrix}
\alpha & -\mu &&&& \\
-\alpha & \alpha + \mu & -\mu &&& \\
& -\alpha & \alpha + \mu & -\mu && \\
&& \ddots & \ddots & \ddots & \\
&&& -\alpha & \alpha + \mu & -\mu \\
&&&& -\alpha & \mu \\
\end{pmatrix}
\eeq
is an $N+1$ by $N+1$ tridiagonal matrix. It is the matrix form of the linear operator $\Lh$. Here, for concreteness we have assumed that the polymer can grow only up to a maximum length $n=N$, but all subsequent results remain valid in the limit $N\to\infty$.
A maximum length could correspond physically to a polymer growing in a spatially confined domain, but here we introduce it simply as a mathematical convenience.

The eigenvalues of a class of tridiagonal matrices, of which \eref{tridiag} is a member, have been derived analytically \cite{yueh2005} using clever methods of manipulating integer sequences \cite{cheng2003}. For \eref{tridiag} the eigenvalues satisfy \cite{yueh2005}
\beq
\elabel{ev_poly}
\lambda_j = \alpha + \mu + 2\sqrt{\alpha\mu}\cos\theta_j,
\eeq
where $\theta_j$ is restricted by
\beqn
0 &=& \alpha\mu\sin(N\theta_j) + \alpha\mu\sin[(N+2)\theta_j] \nonumber \\
\elabel{theta}
&&+\ (\alpha+\mu)\sqrt{\alpha\mu}\sin[(N+1)\theta_j]
\eeqn
and $\theta_j \neq m\pi$ for integer $m$. Using the trigonometric identity $\sin(a+b) = \sin a\cos b+\sin b\cos a$ on the first line of \eref{theta} we obtain
\beqn
0 &=& [2\alpha\mu\cos\theta_j + (\alpha+\mu)\sqrt{\alpha\mu}]\sin[(N+1)\theta_j] \\
\elabel{condition}
&=& \sqrt{\alpha\mu}\lambda_j\sin[(N+1)\theta_j],
\eeqn
where the second step follows from \eref{ev_poly}. For \eref{condition} to be true, we must either have $\lambda_j = 0$ or $(N+1)\theta_j = j\pi$ for any integer $j$. The set of integers $j$ that yield independent values of $\lambda_j$ in \eref{ev_poly} and also satisfy $\theta_j \neq m\pi$ are $j \in \{ 1, 2, \dots, N \}$. Therefore the eigenvalues are
\beq
\elabel{ev_poly2}
\lambda_j = 
\begin{cases}
0 & j = 0 \\
\alpha + \mu - 2\sqrt{\alpha\mu}\cos\left(\frac{j\pi}{N+1}\right) & 1 \le j \le N,
\end{cases}
\eeq
where we have freely changed the sign of the cosine term due to its symmetry with respect to $j$, so that $\lambda_j$ increases with $j$. \eref{ev_poly2} shows that, apart from the zero eigenvalue, the eigenvalues are confined within the region from $\alpha+\mu-2\sqrt{\alpha\mu}$ to $\alpha+\mu+2\sqrt{\alpha\mu}$ (see \fref{eigen}H). Even when we take $N \to \infty$, the range of the eigenvalues remains finite, while their density becomes infinite. This implies that, in contrast to the cases of gene expression and protein modification where there are fast and slow modes, in polymerization there are only slow modes: every eigenfunction (except the stationary mode) relaxes on a timescale that is on the order of $\alpha+\mu$.

With the eigenvalues known, the eigenfunctions are straightforward to compute using the matrix form of the eigenvalue relation, ${\cal L}\vec{\phi}^j = \lambda_j\vec{\phi}^j$. For example, when $j=0$ we know that $\vec{\phi}^0$ is equivalent to the stationary distribution,
\beq
\elabel{phi0}
\phi_n^0 = (1-\gamma)\gamma^n.
\eeq
When $j > 0$, we find $\vec{\phi}^j$ by solving the eigenvalue relation for each row $n = 0, 1, 2, \dots$ in succession. This is equivalent to the iteration and pattern-matching procedure used to find the stationary distribution (see \aref{exercises}). The result is
\beq
\phi_n^j = (1-\gamma) \sum_{\ell=0}^n (-1)^{n+\ell}{\lfloor\frac{n+\ell}{2}\rfloor\choose\ell}
	(-\gamma)^{\lceil\frac{n-\ell-1}{2}\rceil} (\sqrt{\gamma}\chi_j)^\ell
\eeq
for $j>0$, where $\chi_j \equiv 2\cos[j\pi/(N+1)]$. Here we use $\lfloor\cdot\rfloor$ and $\lceil\cdot\rceil$ to denote the floor and ceiling functions, respectively, and we freely choose the prefactor to match that in \eref{phi0}. Several eigenfunctions are shown in \fref{eigen}G. The conjugate eigenfunctions satisfy $\vec{\psi}^j{\cal L} = \lambda_j\vec{\psi}^j$. That is, in matrix notation, the eigenfunctions are column vectors while the conjugate eigenfunctions are row vectors. The conjugate eigenfunctions are similarly found by iteration and pattern-matching. They read
\beq
\elabel{psi_poly}
\psi_n^j =
\begin{cases}
1 & j = 0 \\
\sum_{\ell=0}^n {\lfloor\frac{n+\ell}{2}\rfloor\choose\ell}
	(-\gamma)^{-\lceil\frac{n-\ell}{2}\rceil} \left( \frac{\chi_j}{\sqrt{\gamma}}\right)^\ell & j > 0,
\end{cases}
\eeq
up to a constant prefactor that can be chosen to satisfy orthonormality with $\phi_n^j$. Just as for gene expression and protein modification, the eigenfunctions and conjugate eigenfunctions form a basis in which any distribution $q_n$ can be expanded, as shown in \fref{eigen}I.

\section{Extensions and outlook}
\slabel{extensions}

In this chapter, we have introduced and solved minimal stochastic models of three canonical processes in cell biology: gene expression, protein modification, and polymerization. We have developed a set of analytic tools ranging from straightforward iteration, to the use of the generating function, eigenfunction expansion, and raising and lowering operators from quantum mechanics. These tools allow one to solve a given problem in multiple ways, and they often lead to important physical insights. In particular, we have seen that the eigenvalues tell us about the relaxation dynamics of a stochastic process, and the eigenfunctions form a convenient basis for expansion. In principle, exploiting the eigenfunctions is always possible because the master equation is a linear equation.

These and other tools have been used to study more complex and realistic processes that extend beyond the three minimal models considered here. In the context of gene expression, it is now known that proteins are often not produced one molecule at a time, but instead in quick bursts of several or tens of molecules at a time \cite{golding2005, raj2006, raj2008}. Additionally, the expression levels of different genes' proteins are far from independent. Rather, many genes express proteins called transcription factors that regulate the expression of other genes. These regulatory interactions form networks in which phenotypic, developmental, and behavioral information is encoded \cite{alon2006}. Many researchers have used the generating function, eigenfunction expansion, and operator methods to solve models of bursty gene expression, gene regulation, and gene regulation with bursts \cite{hornos2005, raj2006, shahrezaei2008, iyerbiswas2009, walczak2009, mugler2009, walczak2012, kumar2014}.

Protein modification events also occur in a tightly regulated manner among different protein types. Collectively these coupled modification events form cell signaling networks. Since modification is faster than gene expression, signaling networks often encode cellular responses that need to be temporally and spatially precise, such as rapid behavioral responses to environmental signals \cite{lim2014}. Indeed, operator methods, field theory, and the renormalization group haven proven especially useful in the analysis of spatially heterogeneous signaling processes \cite{mattis1998, lee1995}. Eigenfunction expansion has also been used to study spatially heterogenous protein modification at the cell membrane \cite{mugler2013}. In general, many of the tools that we have presented in this chapter can be extended to a spatially resolved context \cite{vankampen1992, gardiner1985}.

Finally, polymerization can be far more complex than the model considered here. Microtubules undergo periods of steady growth followed by periods of rapid shrinkage, a process termed dynamic instability \cite{boal2012}. Both microtubules and actin filaments actively regulate their length, often via the action of molecular motors, resulting in relative fluctuations that are much smaller than for the geometric distribution (\fref{processes}I) \cite{kuan2013, mohapatra2015}. Straightforward iteration and more sophisticated analytic techniques have been used to solve models of dynamic instability, length regulation, and other complex polymerization processes \cite{hill1984, verde1992, kuan2013, mohapatra2015}.

Fluctuations dominate almost all processes at the scale of the cell. Stochastic models will continue to be necessary to understand how cells suppress or exploit these fluctuations. Going forward, our hope is that these analytic tools will be expanded upon and extended to new problems, allowing minimal models to remain a powerful complement to computer simulations and experiments in understanding cell function.

\appendix

\section{Exercises for the reader}
\alabel{exercises}

\begin{enumerate}
\item
From the stationary state of \eref{me_gene}, derive \eref{poisson} by iteration. That is, set $n=0$ to find $p_1$ in terms of $p_0$, then set $n=1$ to find $p_2$, and so on until a pattern is identified. What sets $p_0$? Repeat for \erefs{me_mod}{me_poly} to derive \erefs{binomial}{geometric}, respectively. Finally, repeat for the $\vec{\phi}^j$ and $\vec{\psi}^j$ eigenfunction relations to derive \erefn{phi0}{psi_poly}. This last task may be aided by knowledge of some integer sequences, e.g.\ from \cite{oeis}.

\item
Derive \eref{pde_gene} by multiplying \eref{me_gene} by $z^n$ and summing both sides over $n$. Hint: distribute the sum over all four terms on the right-hand side, and where necessary shift the index of summation to obtain $p_n$ instead of $p_{n\pm1}$. Repeat for \erefs{me_mod}{me_poly} to derive \erefs{pde_mod}{pde_poly}, respectively.

\item
Derive \eref{phi_gene} from \erefs{ef_gene}{ev_gene} by taking derivatives (see \eref{phidef}). Hint: the $n$th derivative of a product follows a binomial expansion, $\partial_x^n(fg) = \sum_{k=0}^n {n\choose k} (\partial_x^k f) (\partial_x^{n-k} g)$. Repeat for \eref{ef_mod} to derive \eref{phi_mod}.

\item
Calculate the relative fluctuations $\sigma/\avg{n}$ for the binomial (\eref{binomial}) and geometric distributions (\eref{geometric}), writing the expressions entirely in terms of $\avg{n}$ (and $N$ for the binomial distribution). How do the expressions compare to that for the Poisson distribution, $\sigma/\avg{n} = \avg{n}^{-1/2}$? Sketch a plot of $\sigma/\avg{n}$ vs.\ $\avg{n}$ for all three distributions.

\end{enumerate}



\begin{thebibliography}{10}

\bibitem{guptasarma1995}
Purnananda Guptasarma.
\newblock Does replication-induced transcription regulate synthesis of the
  myriad low copy number proteins of escherichia coli?
\newblock {\em Bioessays}, 17(11):987--997, 1995.

\bibitem{elowitz2002}
Michael~B Elowitz, Arnold~J Levine, Eric~D Siggia, and Peter~S Swain.
\newblock Stochastic gene expression in a single cell.
\newblock {\em Science}, 297(5584):1183--1186, 2002.

\bibitem{raj2006}
Arjun Raj, Charles~S Peskin, Daniel Tranchina, Diana~Y Vargas, and Sanjay
  Tyagi.
\newblock Stochastic mrna synthesis in mammalian cells.
\newblock {\em PLoS Biol}, 4(10):e309, 2006.

\bibitem{gillespie1977}
Daniel~T Gillespie.
\newblock Exact stochastic simulation of coupled chemical reactions.
\newblock {\em The journal of physical chemistry}, 81(25):2340--2361, 1977.

\bibitem{erban2007}
Radek Erban, Jonathan Chapman, and Philip Maini.
\newblock A practical guide to stochastic simulations of reaction-diffusion
  processes.
\newblock {\em arXiv:0704.1908}, 2007.

\bibitem{gillespie2001}
Daniel~T Gillespie.
\newblock Approximate accelerated stochastic simulation of chemically reacting
  systems.
\newblock {\em The Journal of Chemical Physics}, 115(4):1716--1733, 2001.

\bibitem{rathinam2003}
Muruhan Rathinam, Linda~R Petzold, Yang Cao, and Daniel~T Gillespie.
\newblock Stiffness in stochastic chemically reacting systems: The implicit
  tau-leaping method.
\newblock {\em The Journal of Chemical Physics}, 119(24):12784--12794, 2003.

\bibitem{hornos2005}
JEM Hornos, D~Schultz, GCP Innocentini, JAMW Wang, AM~Walczak, JN~Onuchic, and
  PG~Wolynes.
\newblock Self-regulating gene: an exact solution.
\newblock {\em Physical Review E}, 72(5):051907, 2005.

\bibitem{shahrezaei2008}
Vahid Shahrezaei and Peter~S Swain.
\newblock Analytical distributions for stochastic gene expression.
\newblock {\em Proceedings of the National Academy of Sciences},
  105(45):17256--17261, 2008.

\bibitem{iyerbiswas2009}
Srividya Iyer-Biswas, F~Hayot, and C~Jayaprakash.
\newblock Stochasticity of gene products from transcriptional pulsing.
\newblock {\em Physical Review E}, 79(3):031911, 2009.

\bibitem{walczak2009}
Aleksandra~M Walczak, Andrew Mugler, and Chris~H Wiggins.
\newblock A stochastic spectral analysis of transcriptional regulatory
  cascades.
\newblock {\em Proceedings of the National Academy of Sciences},
  106(16):6529--6534, 2009.

\bibitem{mugler2009}
Andrew Mugler, Aleksandra~M Walczak, and Chris~H Wiggins.
\newblock Spectral solutions to stochastic models of gene expression with
  bursts and regulation.
\newblock {\em Physical Review E}, 80(4):041921, 2009.

\bibitem{kumar2014}
Niraj Kumar, Thierry Platini, and Rahul~V Kulkarni.
\newblock Exact distributions for stochastic gene expression models with
  bursting and feedback.
\newblock {\em Physical review letters}, 113(26):268105, 2014.

\bibitem{hespanha2005}
Joao~Pedro Hespanha and Abhyudai Singh.
\newblock Stochastic models for chemically reacting systems using polynomial
  stochastic hybrid systems.
\newblock {\em International Journal of robust and nonlinear control},
  15(15):669--689, 2005.

\bibitem{munsky2006}
Brian Munsky and Mustafa Khammash.
\newblock The finite state projection algorithm for the solution of the
  chemical master equation.
\newblock {\em The Journal of chemical physics}, 124(4):044104, 2006.

\bibitem{vankampen1992}
Nicolaas~Godfried Van~Kampen.
\newblock {\em Stochastic processes in physics and chemistry}, volume~1.
\newblock Elsevier, 1992.

\bibitem{gardiner1985}
Crispin~W Gardiner.
\newblock {\em Handbook of stochastic methods}, volume~4.
\newblock Springer Berlin, 1985.

\bibitem{doi1976}
Masao Doi.
\newblock Second quantization representation for classical many-particle
  system.
\newblock {\em Journal of Physics A: Mathematical and General}, 9(9):1465,
  1976.

\bibitem{zeldovich1978}
Ya~B Zel'Dovich and AA~Ovchinnikov.
\newblock The mass action law and the kinetics of chemical reactions with
  allowance for thermodynamic fluctuations of the density.
\newblock {\em Zh. Eksp. Teor. Fiz}, 74:1588--1598, 1978.

\bibitem{peliti1986}
L~Peliti.
\newblock Renormalisation of fluctuation effects in the a+ a to a reaction.
\newblock {\em Journal of Physics A: Mathematical and General}, 19(6):L365,
  1986.

\bibitem{mattis1998}
Daniel~C Mattis and M~Lawrence Glasser.
\newblock The uses of quantum field theory in diffusion-limited reactions.
\newblock {\em Reviews of Modern Physics}, 70(3):979, 1998.

\bibitem{walczak2012}
Aleksandra~M Walczak, Andrew Mugler, and Chris~H Wiggins.
\newblock Analytic methods for modeling stochastic regulatory networks.
\newblock In {\em Computational Modeling of Signaling Networks}, pages
  273--322. Springer, 2012.

\bibitem{mugler2013}
Andrew Mugler, Filipe Tostevin, and Pieter~Rein ten Wolde.
\newblock Spatial partitioning improves the reliability of biochemical
  signaling.
\newblock {\em Proceedings of the National Academy of Sciences},
  110(15):5927--5932, 2013.

\bibitem{alberts2007}
Bruce Alberts, Alexander Johnson, Julian Lewis, Martin Raff, Keith Roberts, and
  Peter Walter.
\newblock {\em Molecular biology of the cell}.
\newblock Garland Science, 2007.

\bibitem{phillips2012}
Rob Phillips, Jane Kondev, Julie Theriot, and Hernan Garcia.
\newblock {\em Physical biology of the cell}.
\newblock Garland Science, 2012.

\bibitem{alon2006}
Uri Alon.
\newblock {\em An introduction to systems biology: design principles of
  biological circuits}.
\newblock CRC press, 2006.

\bibitem{mugler_thesis}
Andrew Mugler.
\newblock {\em Form and function in small biological networks}.
\newblock PhD thesis, Columbia University, 2010.

\bibitem{townsend2000}
John~S Townsend.
\newblock {\em A modern approach to quantum mechanics}.
\newblock University Science Books, 2000.

\bibitem{lim2014}
Wendell Lim, Bruce Mayer, and Tony Pawson.
\newblock {\em Cell Signaling: principles and mechanisms}.
\newblock Taylor \& Francis, 2014.

\bibitem{boal2012}
David Boal.
\newblock {\em Mechanics of the Cell}.
\newblock Cambridge University Press, 2012.

\bibitem{yueh2005}
Wen-Chyuan Yueh.
\newblock Eigenvalues of several tridiagonal matrices.
\newblock {\em Applied Mathematics E-Notes}, 5(66-74):210--230, 2005.

\bibitem{cheng2003}
Sui~Sun Cheng.
\newblock {\em Partial difference equations}, volume~3.
\newblock CRC Press, 2003.

\bibitem{golding2005}
Ido Golding, Johan Paulsson, Scott~M Zawilski, and Edward~C Cox.
\newblock Real-time kinetics of gene activity in individual bacteria.
\newblock {\em Cell}, 123(6):1025--1036, 2005.

\bibitem{raj2008}
Arjun Raj and Alexander van Oudenaarden.
\newblock Nature, nurture, or chance: stochastic gene expression and its
  consequences.
\newblock {\em Cell}, 135(2):216--226, 2008.

\bibitem{lee1995}
Benjamin~P Lee and John Cardy.
\newblock Renormalization group study of thea+ b?? diffusion-limited reaction.
\newblock {\em Journal of statistical physics}, 80(5-6):971--1007, 1995.

\bibitem{kuan2013}
Hui-Shun Kuan and MD~Betterton.
\newblock Biophysics of filament length regulation by molecular motors.
\newblock {\em Physical biology}, 10(3):036004, 2013.

\bibitem{mohapatra2015}
Mohapatra L, Goode BL, and Kondev J.
\newblock Antenna mechanism of length control of actin cables.
\newblock {\em PLoS Comput Biol}, 11(6):e1004160, 2015.

\bibitem{hill1984}
Terrell~L Hill.
\newblock Introductory analysis of the gtp-cap phase-change kinetics at the end
  of a microtubule.
\newblock {\em Proceedings of the National Academy of Sciences},
  81(21):6728--6732, 1984.

\bibitem{verde1992}
Fulvia Verde, Marileen Dogterom, Ernst Stelzer, Eric Karsenti, and Stanislas
  Leibler.
\newblock Control of microtubule dynamics and length by cyclin a-and cyclin
  b-dependent kinases in xenopus egg extracts.
\newblock {\em The Journal of Cell Biology}, 118(5):1097--1108, 1992.

\bibitem{oeis}
The online encyclopedia of integer sequences.
\newblock \url{http://oeis.org}.

\end{thebibliography}

\end{document}